\documentclass[epj]{svjour}
\usepackage{graphicx}
\usepackage{amsmath}
\usepackage{amssymb}
\usepackage{verbatim}

\begin{document}

\title{Cartesian and polar Schmidt bases for down-converted photons}
\subtitle{How high dimensional entanglement protects the shared information from non-ideal measurements}
\author{Filippo M. Miatto, Thomas Brougham, Alison M. Yao}
\institute{SUPA and Department of Physics, University of Strathclyde, Glasgow G4 0NG, Scotland, U.K.}
\date{\today}
\authorrunning{Miatto et al.}
\titlerunning{Cartesian and polar Schmidt bases for SPDC photons}
\abstract{
We derive an analytical form of the Schmidt modes of spontaneous parametric down-conversion (SPDC) biphotons in both cartesian and polar coordinates. We show that these correspond to Hermite-Gauss (HG) or Laguerre-Gauss (LG) modes only for a specific value of their width, and we show how such value depends on the experimental parameters.
The Schmidt modes that we explicitly derive allow one to set up an optimised projection basis that maximises the mutual information gained from a joint measurement. The possibility of doing so with LG modes makes it possible to take advantage of the properties of orbital angular momentum eigenmodes.
We derive a general entropic entanglement measure using the R\'enyi entropy as a function of the Schmidt number, $K$, and then retrieve the von Neumann entropy, $S$. Using the relation between $S$ and $K$ we show that, for highly entangled states, a non-ideal measurement basis does not degrade the number of shared bits by a large extent. More specifically, given a non-ideal measurement which corresponds to the loss of a fraction of the total number of modes, we can quantify the experimental parameters needed to generate an entangled SPDC state with a sufficiently high dimensionality to retain any given fraction of shared bits.}
\maketitle

\section{Introduction}

Entangled states are one of the most striking predictions of quantum mechanics. Their generation and 
measurement not only allows for tests of the theory but may also result in novel technological advances. 
Entangled states form the basis of quantum information and quantum computing and are essential for 
applications such as quantum cryptography where they are used for distributing secret cryptographic 
keys between two parties \cite{qkd,qkd2,qkd3}.

Quantum systems can be entangled in various degrees of freedom, e.g. polarisation, conjugate variables 
such as time and energy, position and momentum, and also in multiple variables simultaneously 
(known as hyperentanglement) \cite{hyper1,hyper2}. Recently there has been a lot of interest in demonstrating 
entanglement between spatial modes carrying orbital angular momentum (OAM) \cite{Jack10}. These 
are of particular interest for quantum information protocols, such as quantum key distribution (QKD), 
as they reside within an infinite-dimensional, discrete Hilbert space, thus allowing large amounts of information 
to be impressed onto a single photon and multiple bits of secret key to be extracted for each entangled state.

Experimentally, pairs of photons entangled in their OAM can be reliably produced using spontaneous parametric 
down-conversion (SPDC) \cite{BoydNLOptics} and there has been a lot of effort to calculate the precise form of the 
down-converted photons, and hence quantify the ensuing degree of entanglement, by calculating the 
\textit{spiral bandwidth} of the system \cite{Torres03,Miatto11,Yao11}. Alternatively, the amount of entanglement can
be quantified by performing a Schmidt decomposition of the down-converted state \cite{barnett}. This gives information
both about the pairing of the photons and about the degree of entanglement of the state via the Schmidt number, $K$, 
which is defined as the average number of non-zero coefficients in the Schmidt decomposition. For an entangled bipartite 
pure state the maximum possible correlation will occur when the measurements are performed in states which correspond to 
a Schmidt decomposition of the bipartite state \cite{indexcor,hall,hallcorr}. 

Modes carrying OAM are most commonly described in polar coordinates by superpositions of Laguerre-Gaussian 
(LG) modes but may, equivalently, be described in cartesian coordinates by superpositions of Hermite-Gaussian (HG) modes.
In this paper we calculate explicitly the Schmidt bases for both coordinate systems of the SPDC state under the assumption of gaussian phase matching. Such bases can be experimentally implemented for joint measurements. We use our results to calculate the Schmidt number of the state and its R\'enyi entropy. This not only gives an alternative measure of the entanglement, but also allows us, as a special case, to calculate the von Neumann entropy of these reduced states, and hence the maximum information that can be shared as a function of the experimental parameters.

The paper is organised as follows. In section 2 we describe the biphoton state produced using SPDC. We show its dependence on the pump parameters and crystal length and describe the gaussian phase matching approximation that can be made in the near collinear limit. This allows us, in section 3, to calculate the Schmidt decomposition in both cartesian and polar coordinates, and to show how one can easily transform between the two. In section 4 we use this result to analyse the entanglement by means of the Schmidt number and hence the R\'enyi and von Neumann entropies. We then demonstrate the simple relation between the information entropy and the Schmidt number and show how the information loss due to non-ideal measurements depends on the number of entangled modes and how high dimensional entanglement protects the shared information.

\section{The SPDC state}

In spontaneous parametric down-conversion (SPDC) two lower-frequency photons, commonly referred to as signal 
and idler, are generated when a pump field interacts with a nonlinear crystal \cite{Hong85}. The spatial structure of 
the down-converted biphotons depends both on the pump field and on the phase matching. For a gaussian pump it 
can be written \cite{carlosgauss}:
\begin{align}
\psi(\mathbf{q}_i,\mathbf{q}_s)=\mathcal{N}\exp\left(-\frac{w_p^2}{4}|\mathbf{q}_s+\mathbf{q}_i|^2\right)\mathrm{sinc}\left(\frac{L \Delta k_z}{2} +\Phi\right)
\end{align}
where $\mathbf{q}_{s,i}$ are the transverse components of the wave vectors $\mathbf{k}_{s,i}$ for the signal and 
idler fields, $\Delta k_z$ is the longitudinal component of the wave mismatch between the pump wave vector $\mathbf{k}_p$ 
and the down-converted photons wave vectors: 
$\Delta k_z=(\mathbf{k}_{p}-\mathbf{k}_{i}-\mathbf{k}_{s})_z\simeq \left(|\mathbf{q}_i-\mathbf{q}_s|^2\right)/k_p$, 
where we have made use of the paraxial approximation. $\Phi$ is an additional phase mismatch which depends on the 
internal refractive indices (i.e. its value can be tuned by tilting the crystal or by changing its temperature). For our 
analysis $\Phi=0$.

Near the collinear phase matching regime the analysis can be simplified by using a gaussian approximation of the phase matching term \cite{laweberly,carlosgauss} so the state 
can be written in the form:
\begin{align}
\psi(\mathbf{q}_i,\mathbf{q}_s) \propto \exp\left(-\frac{|\mathbf{q}_i+\mathbf{q}_s|^2}{\sigma^2}\right)\exp\left(-b^2|\mathbf{q}_i-\mathbf{q}_s|^2\right)
\label{spdcstate}
\end{align}
where $b$ and $\sigma$ depend on the pump waist, $w_p$, and wave number, $k_p$, and on the crystal 
length, $L$, in the following way
\begin{equation}
b=\frac{1}{2} \sqrt{\frac{L}{k_p}}\, ; \, \sigma=\frac{2}{w_p}.
\end{equation}
If we scale the wavevectors $\mathbf{q}_{i,s}$ by a factor $\Gamma' = \sqrt{b/ \sigma}$ s.t. $\mathbf{q} = \mathbf{q}'/\Gamma'$ then we can re-write
(\ref{spdcstate}) as
\begin{align}
\psi(\mathbf{q'}_i,\mathbf{q'}_s) \propto \exp\left(-\frac{|\mathbf{q'}_i+\mathbf{q'}_s|^2}{b \sigma}\right)
\exp\left(-b \sigma |\mathbf{q'}_i-\mathbf{q'}_s|^2\right)
\label{spdcstate2}
\end{align}
where
\begin{align}
b\sigma= \sqrt{\frac{L}{2z_r}}
\label{bsigma}
\end{align}
with $z_r$ the Rayleigh range of the pump beam. Writing (\ref{spdcstate}) in this form allows us to see the 
symmetry between the two parts of the wave function and will help explain some of our later results. In
particular, we can see that our results will depend on the product $b \sigma$ and not $b$ and $\sigma$ 
independently.

\section{Schmidt decomposition}

As we stated earlier, a useful, and experimentally convenient, measure of the entanglement is the
Schmidt number, $K$. In order to calculate this we first need to calculate the Schmidt decomposition
of the down-converted state, which is done by writing it in the form 
\begin{equation}
\label{defschmidt}
|\Psi\rangle_{AB}=\sum_i{\sqrt{\lambda_i}|\alpha_i\rangle_A|\beta_i\rangle_B} ,
\end{equation}
where $|\alpha_i\rangle_A, |\beta_i\rangle_B$ are the Schmidt modes, defined by the eigenvectors of the 
reduced density matrices, and the real and positive Schmidt coefficients, $\sqrt{\lambda_i}$, are the corresponding eigenvalues,
with each of the factors in the normalized set $\{\lambda_i\}$ representing the probability of detecting 
the entangled state in the $i^{\mathrm{th}}$ entangled Schmidt mode $|\alpha_i\rangle_A|\beta_i\rangle_B$.
If all of the coefficients $\sqrt{\lambda_i}$ are different, then the Schmidt decomposition is unique.
Whenever some of the coefficients $\sqrt{\lambda_i}$ are equal, one has a choice of infinitely 
many different Schmidt bases.

The Schmidt decomposition provides insights into the nature of the bipartite entanglement by determining 
the natural set of biorthogonal mode pairs (or orthonormal bases) for the two systems 
\cite{barnett,laweberly,nch,lawschmidt,ekart} while the coefficients allow us to calculate the Schmidt number, 
$K$  (i.e. the average number of modes in the state) and the entropy of entanglement. By knowing the Schmidt decomposition explicitly in HG and LG modes, one can easily implement an ideal detection basis.

In this paper we are interested in the biphoton state produced using SPDC which is entangled in its
OAM.  As this can be equivalently described using either HG or LG modes we perform the Schmidt 
decomposition in the two corresponding coordinate systems. By approximating the biphoton state 
as the double Gaussian in (\ref{spdcstate}) we are able to calculate the Schmidt decomposition in an 
analytical form for both. We also demonstrate their equivalence by using the well-known relationship 
between HG and LG modes \cite{hgtolg}.
We then show that, as expected, the entropic analysis leads to the same conclusions for both.
Note that in both cases, the Schmidt modes for the SPDC state will have the same form for both signal and 
idler, due to symmetry requirements \cite{lawschmidt} and so we will obtain a decomposition of the 
form:
\begin{align}
\psi(\mathbf{q}_i,\mathbf{q}_s)=\sum_{a,b}\sqrt{\lambda_{a,b}}\,u_{a,b}(\mathbf{q}_i)u^*_{a,b}(\mathbf{q}_s) ,
\label{decomp}
\end{align}
where the functions $u_{a,b}$ depend on the coordinate system employed, and the labels $a$ and $b$ 
correspond to different degrees of freedom: in the cartesian case $a$ and $b$ will be replaced with $m$ and $n$, in the polar case with $\ell$ and $p$. The $m$ and $n$ quantum numbers label the two transverse degrees of freedom, while the $\ell$ and $p$ quantum numbers label the angular 
and radial degrees of freedom, respectively. The sum is calculated on two indices because we perform the decomposition in the two-dimensional plane perpendicular to the direction of emission.

\subsection{Decomposition in cartesian coordinates}

In a recent paper, Straupe \textit{et al.} \cite{Straupe11} reported a proof-of-principle experiment demonstrating that 
an appropriately chosen set of HG modes constitutes a Schmidt decomposition for transverse momentum states of 
biphotons generated by SPDC. For clarity and completeness we perform an equivalent Schmidt decomposition 
in cartesian coordinates (details of the calculations are given in Appendix \ref{sec-AppA}) before extending our analysis 
to polar coordinates in the next section and then demonstrating their equivalence.

The cartesian decomposition requires a separation of each of the variables $\mathbf{q}_i$ and $\mathbf{q}_s$ into a pair of orthogonal variables, $q$ and $q_\bot$, so that the wave function assumes the form $\psi(\mathbf{q}_i,\mathbf{q}_s)\rightarrow\psi(q_i,q_s,q_{i\bot},q_{s\bot})$.
We define the cartesian basis of HG modes as
\begin{align}
h_n(\Gamma q)=\frac{\sqrt{\Gamma}e^{-\Gamma^2q^2/2}H_n(\Gamma q)}{(n!2^n\sqrt{\pi})^{1/2}}
\label{cartbasis}
\end{align}
where $ \Gamma = 2 \sqrt{\frac{b}{\sigma}}=w_p\sqrt[4]{\frac{L}{2z_r}}$ is the width of the HG modes. If we express the wave 
function (\ref{spdcstate}) in terms of this basis we obtain
\begin{equation}
\label{schm}
\psi=(1-\mu^2)\sum_{m,n\geq 0}\mu^{m+n}h_{mn}(q_i,q_{i\bot})h_{mn}(q_s,q_{s\bot}) ,
\end{equation}
where  
\begin{equation}
h_{mn}(x,y)=h_m(\Gamma x)h_n(\Gamma y)
\end{equation} 
and 
\begin{equation}
\mu=\left |\frac{b\sigma-1}{b\sigma+1}\right|.
\label{eqn-mu}
\end{equation}

Expression \eqref{schm} is the cartesian Schmidt form for the SPDC state with corresponding coefficients:
\begin{align}
\sqrt{\lambda_{m,n}}=(1-\mu^2)\mu^{m+n} = \frac{4 b \sigma}{\left( 1+b \sigma \right)^2} \left| \frac{b \sigma - 1}{b \sigma + 1} \right|^{m+n} .
\label{schmidtcoeffcartesian}
\end{align}
Note that these are exactly equivalent to those given in equation (5) of \cite{Straupe11}.

\subsection{Decomposition in polar coordinates}

As LG modes are currently the preferred basis for many spatial entanglement experiments we also 
calculate the Schmidt decomposition in polar coordinates (details of the calculations are given in 
Appendix \ref{sec-AppB}). The polar decomposition requires a separation of each of the variables $\mathbf{q}_i$ 
and $\mathbf{q}_s$ into a pair of polar variables, so that the wave function assumes the form 
$\psi(\mathbf{q}_i,\mathbf{q}_s)\rightarrow\psi(\rho_i,\rho_s,\varphi_{i},\varphi_{s})$, where $\rho$ and 
$\varphi$ are the radial and angular variables.
We take the LG modes to have the standard definition in momentum space, and we add the same scaling factor $\Gamma$ as in the cartesian case: 
\begin{align}
LG^{\ell}_p(\Gamma\rho,\varphi)=\sqrt{\frac{\Gamma^2p!}{\pi(p+|\ell|)!}}e^{-\frac{\Gamma^2\rho^2}{2}}\left(\Gamma\rho \right)^{|\ell|}L_p^{(|\ell|)}\left(\Gamma\rho^2\right)e^{i\ell\varphi}
\label{polbasis}
\end{align}
where $L_p^{(\ell)}$ are generalized Laguerre polynomials.  One can express the wave function (\ref{spdcstate}) in terms of LG modes, of width $\Gamma/\sqrt2$ 
\begin{align}
\psi=(1-\mu^2)\sum_{\ell=-\infty}^\infty\sum_{p=0}^\infty \mu^{2p+|\ell|}LG^{\ell}_p(\Gamma\rho_i,\varphi_i)LG^{-\ell}_p(\Gamma\rho_s,\varphi_s).
\label{pol}
\end{align}
where $\Gamma = 2 \sqrt{\frac{b}{\sigma}}=w_p\sqrt[4]{\frac{L}{2z_r}}$.

The above expression is the polar coordinate form for the Schmidt decomposition of the SPDC state, where $\mu$ is defined in (\ref{eqn-mu})
and the polar Schmidt coefficients are given by
\begin{align}
\sqrt{\lambda_{\ell,p}}=(1-\mu^2)\mu^{2p+|\ell|}= \frac{4 b \sigma}{\left( 1+b \sigma \right)^2} \left| \frac{b \sigma - 1}{b \sigma + 1} \right|^{2p+|\ell|} ,
\label{schmidtcoeffpolar}
\end{align}
to be compared to (\ref{schmidtcoeffcartesian}).

\subsection{Equivalence of Schmidt bases}
The expressions calculated above are equivalent descriptions of the entangled state and, just as it is possible
to transform LG modes into HG modes \cite{Allen} and vice versa \cite{hgtolg}, we are also able to
convert between our two Schmidt bases.
In fact, it is straightforward to convert the Schmidt decomposition in cartesian coordinates, equation (\ref{cartbasis}), 
into that in polar coordinates, (\ref{polbasis}).  The first step is to notice that the values of 
$m$ and $n$ that satisfy $m+n=N$ yield the same Schmidt coefficient $(1-\mu^2)\mu^{N}$; for the polar 
case this happens for all the values of $\ell$ and $p$ that satisfy $|\ell|+2p=N$. The number $N$ is called the 
mode order.

The relation between the cartesian basis \eqref{cartbasis} and the polar basis \eqref{polbasis} is \cite{hgtolg}
\begin{align}
LG_p^{(\ell)}(\rho)=\sum_{k=0}^Nb_{p,k}^{(N)}h_{N-k,k}(q,q_\bot)
\end{align}
where the relation between cartesian and polar coordinates is the canonical one, and
\begin{align}
b_{p,k}^{(N)}=\frac{i^k(-1)^{p+k}}{2^{N/2}k!}\frac{d^k}{d\mu^k}[(1-\mu)^n(1+\mu)^m]_{t=0}
\end{align}
where $m+n=|\ell|+2p=N$. This fact enforces the conservation of the mode order when changing Schmidt 
basis.

\section{Analysis of the entanglement}

\subsection{Schmidt number}
The entanglement of a state can be quantified by the probability distribution of the modes it contains. Intuitively, 
a state is more entangled whenever this probability distribution is more `spread out'. A particularly
important measure of entanglement is the Schmidt number, $K$, which corresponds to the number of significant 
modes in the Schmidt decomposition \cite{lawschmidt,ekart}. For states in the form (\ref{decomp}), this is defined as
\begin{align}
K=\frac{\text{Tr}[\hat\rho]^2}{\text{Tr}[\hat\rho^2]} \equiv \frac{1}{\sum_{a,b} \lambda_{a,b}^2}
\end{align}
where $\hat\rho$ is the reduced state formed by tracing over one part of a pure bipartite state and $\sqrt{\lambda_{a,b}}$ 
are the Schmidt coefficients that appear in the Schmidt decomposition of the bipartite state. One can immediately see 
that the eigenvalues of the reduced state are just the square of the Schmidt coefficients.  A state will be separable when 
$K=1$ and entangled if $K>1$. Applying this to the Schmidt decompositions calculated earlier, gives
\begin{align}
K&=\left[\sum_{m=0}^\infty\sum_{n=0}^\infty\lambda_{m,n}^2\right]^{-1} =\left[\sum_{\ell=-\infty}^\infty\sum_{p=0}^\infty\lambda_{\ell,p}^2\right]^{-1} \nonumber \\
&=\frac{1}{4}\left(b\sigma+\frac{1}{b\sigma}\right)^2 .
\label{schmidtnumber}
\end{align}
This result agrees with previous calculations of $K$ \cite{laweberly} and is independent of the Schmidt basis used. This means that one has the freedom to choose the basis that best matches the experimental conditions with no consequence on the dimensionality of the Hilbert space that is spanned by the detection basis.

The effect of the experimental parameters can be seen more clearly if, as in (\ref{bsigma}), we write 
$b\sigma= w_p\sqrt{L/k_p}=\sqrt{L/2z_r}$ where $L$ is the crystal thickness, $w_p$ and $k_p$ are the width 
and wave number, respectively, of the pump and $z_r$ is its Rayleigh range.
Note that $K = 1$, which means that the state is not entangled, whenever $b \sigma = 1$, which 
corresponds to choosing experimental parameters such that the crystal length is twice the Rayleigh
range ($L = 2 z_r$).

\subsection{R\'enyi entropy}

An alternative approach to quantifying the entanglement of an SPDC state is to calculate its entropy.  
The most famous entropic function is the Shannon entropy, which appears in information theory and statistical 
mechanics \cite{shannon,jaynes}. A more general measure, however, is the R\'enyi entropy, which is obtained 
by neglecting the grouping property of entropy \cite{EIT,renyi}.  For a probability distribution $\{p_k\}$  
the R\'eyni entropy of order $\alpha$ is defined as
\begin{equation}
H_\alpha (\{p_k\})=\frac{1}{1-\alpha}\log_2\left(\sum_k{p^{\alpha}_k}\right),\;\;\alpha>0.
\label{rent}
\end{equation}
Note that when $\alpha \rightarrow 1$ one regains the Shannon entropy.

A simple calculation shows that the R\'enyi entropy of the Schmidt coefficients in either eq. \eqref{schmidtcoeffcartesian} 
or \eqref{schmidtcoeffpolar} is
\begin{align}
H_\alpha(b\sigma)=\frac{2}{\alpha-1}\log_2\frac{(4b\sigma)^\alpha}{|1+b\sigma|^{2\alpha}-|1-b\sigma|^{2\alpha}}.
\label{spdcrenyi}
 \end{align}
Using equation (\ref{schmidtnumber}) we can rewrite the parameter $b \sigma$ in terms of the Schmidt
number, $K$, as $b\sigma=\sqrt{K}-\sqrt{K-1}$ and hence find the R\'enyi entropy in terms of the Schmidt 
number. Replacing $b\sigma$ with $\sqrt{K}-\sqrt{K-1}$ in (\ref{spdcrenyi}) gives the R\'enyi entropy as a function of $K$, which can be approximated by
 \begin{align}
H_\alpha(K)\simeq\log_2K-f(\alpha)
 \end{align}
 where $f(\alpha)=2-\log_2(\alpha^2)/(\alpha-1)$. 
For a discussion of the quality of this approximation, see the next subsection.
This tells us that, to a good approximation, valid for sufficiently high $K$, different orders of the R\'enyi entropy differ by a constant value.
Note that $f(2)=0$ and thus $H_2=\log_2 K$, in fact the Schmidt number is related to the R\'enyi entropy of order 2 by $K=2^{H_2}$.

\subsection{Von Neumann entropy}
For quantum systems we can write the quantum R\'enyi entropy as
\begin{equation}
H_\alpha [\hat\rho] = \frac{1}{1-\alpha} \log_2 \left[ \mathrm{Tr} \left( \hat \rho^{\alpha} \right) \right],\;\;\alpha>0 ,
\label{qrent}
\end{equation}
where $\hat \rho$ is the reduced density matrix \cite{FlammiaPRL09}.
An important special case of the R\'enyi entropy is when one takes the limit $\alpha \rightarrow 1$ in which case \eqref{qrent} 
reduces to the von Neumann entropy of the reduced state \cite{barnett,nch}
\begin{equation}
S[\hat\rho]=-\text{Tr}[\hat\rho\log_2\hat\rho].
\end{equation}
The entropy of a reduced state is known as either the index of correlation \cite{indexcor} or the entanglement 
entropy \cite{BDSW,BBPS} and the importance of the entanglement entropy stems from the fact that it quantifies 
the number of entangled bits (or ebits) within the state \cite{barnett,nch,BBPS}.  This means that if one has $n$ copies of an entangled pure state, with entanglement entropy $S$, then one can asymptotically convert this to 
approximately $nS$ maximally entangled states. It has been shown that the maximum amount of shared information that two parties can extract from an entangled 
pure state is given by entanglement entropy of their state \cite{indexcor,indexcor2,hall,hallcorr}.  Our results thus enable us to 
determine the maximum amount of shared bits per photon pair that two parties can extract from SPDC states. 
Using our results, one can also see how changing the parameters of the pump or the crystal affects the amount of shared information.

\begin{figure}[ht]
\center{\includegraphics[width= 0.48 \textwidth]{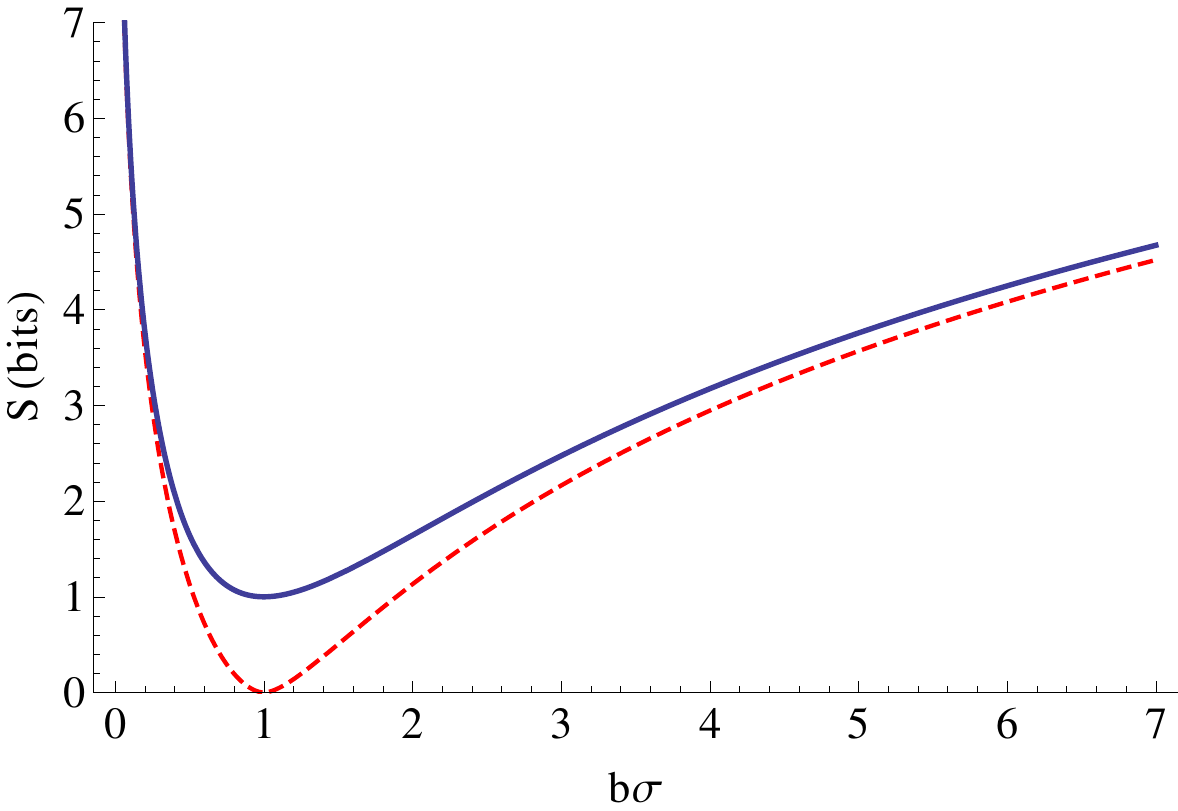}
\caption{A plot of the shared bits of information per photon pair, given in eq. \eqref{vonneuman}, plotted against $b\sigma=\sqrt{L / 2z_r}$ (solid line) and of the non approximated Von Neumann entropy (dashed line). Note that they match for large values of $K$, that is for small or very large values of $b\sigma$.}
\label{figinfo}}
\end{figure}

A straightforward calculation shows that the entanglement entropy for the SPDC state can be approximated by a logarithmic relation:
\begin{align}
S(K)=\lim_{\alpha\rightarrow1}H_\alpha\simeq 1+\log_2(K) .
\label{vonneuman}
\end{align}
Such relation is an approximation that holds well for large enough values of $K$. In fact, a Taylor expansion of the non approximated Von Neumann entropy for large $K$  yields
\begin{align}
S(K)= \underbrace{\frac{2}{\log 2}-2}_{\sim0.9}+\log_2 K-\frac{1}{K\log8}+O\left(\frac{1}{K}\right)^{2}
\label{fullvonneuman}
\end{align}
Without considering such power expansion, it is not immediately obvious how much the relation between Schmidt number and Von Neumann entropy differs from a purely logarithmic approximation. As an example of the failure of the approximation \eqref{vonneuman}, note that in the regime where the state is not entangled, i.e. if there is only one joint mode (and so $K=1$) the amount of quantum correlation in the state has to be 0 (dashed line in Fig. \ref{figinfo}) and not 1 (solid line in Fig. \ref{figinfo}). Common experimental conditions where one seeks high dimensional entanglement are in the range of $b\sigma\ll1$, where the relation \eqref{vonneuman} is accurate. However, experiments with a tightly focussed pump, or a long crystal, may fall in the region closer to $b\sigma\sim1$, where it fails. A plot of the Von Neuman entropy and of its approximation are given in figure \ref{figinfo}.

An interesting feature of figure \ref{figinfo} is that the results are symmetric under the substitution $b\sigma \rightarrow 1/b\sigma$. 
This can be explained by reference to equation (\ref{spdcstate}2) which describes the correlations in the two conjugate planes ($\mathbf{q'}_i+\mathbf{q'}_s$) and ($\mathbf{q'}_i-\mathbf{q'}_s$). Interchanging $b\sigma$ and $1/b\sigma$ corresponds to `squeezing' in one plane but 'expanding' in the other: the product of the two is constant.
Physically this means that an experiment with a crystal of length $L$ and Rayleigh range $z_r$ is equivalent to an experiment 
with crystal length $az_r$ and Rayleigh range of $L/a$. We expect this symmetry to be no longer exact outside the approximation of the 
gaussian phase matching function.

As we already mentioned, the value of the shared bits reaches its minimum for $b\sigma=1$. In this regime the state is separable, i.e. not entangled. In fact, for such value of $b\sigma$, the Schmidt number $K$ is also 1, which means that the state can be written using only one Schmidt mode, which is separable by definition.
In the rest of the paper we will assume a large enough Schmidt number to safely use the definition \eqref{vonneuman} of the entanglement entropy.

The logarithmic dependence of $S(K)$, for large $K$, has two important consequences. Both are due to the small value of the slope of the logarithm curve for large values of the argument. The first is that 
if we can prepare an SPDC state with a large number of modes, and thus large Schmidt number, any further increase in the number of modes will provide only a modest increase in both the entanglement and the number of shared bits that one can extract. The second consequence is concerned with the non-ideal detection of the entangled state and is discussed in the following subsection.

It is interesting to compare this result with the work \cite{FedorovSpace}, in particular the isotropic case, but also \cite{FedorovTime}, where a relation between spectral entanglement and a control parameter similar to \eqref{schmidtnumber} is found. In particular we note that in their assessment of the spectral-temporal part of the down-converted state, the entanglement strength has its minimum at a value considerably larger than zero, meaning that spectral entanglement is never small. In our case, however, the state is spatially separable when $b\sigma = 1$. In both cases, the control parameters depend upon the characteristics of the crystal and of the pump beam and this could have an implication on the extent of hyperentanglement.
\subsection{Non-ideal detection}
There can be many sources of non-ideal detection. These can range from defects in the measuring apparatus (which give rise to cross-talk between channels, accidental coincidences, dark counts, etc.) to turbulence (that can affect the propagation of the states), to non-ideal choices of the optical elements in the setup (which determine a mismatch between the Schmidt modes and the detection modes and therefore impairs the ability to detect high order modes). We concentrate, in particular, on cases in which the number of modes that a measurement apparatus can detect is less than the number of modes that the source is producing. This type of experimental inaccuracy is fundamentally different from the others, which have been studied for instance in \cite{ThomasPRA2012,Ma2007}, the difference being that in the previous work it was always assumed that a detector could have access to all the modes that are produced by the source.
Detection of entangled states by projection onto modes that match the Schmidt modes, such as the ones given by eqs. \eqref{schmidtcoeffcartesian} and \eqref{schmidtcoeffpolar}, will yield the maximum amount of shared bits. However, if the detection basis does not exactly match the Schmidt basis, the effective number of Schmidt modes that are measured, $K_{\mathrm{eff}}$, will be less than $K$.  Provided the state is highly entangled (i.e. large $K$), the logarithmic relationship between the number of shared bits and the Schmidt number, given in equation (\ref{vonneuman}), means that even if the fraction of entangled modes that are detected, $\eta=K_{\mathrm{eff}}/K$, is small, this need not be too detrimental to the fraction of shared bits $S(\eta K)/S(K)$ that one can extract. This result may seem counter-intuitive, however, the key point to note is that information is measured by entropy, not by the number of modes \cite{shannon,EIT}. For example, the number of different messages that one could encode using 4 modes can be described using 2 binary digits, while 8 modes require 3 binary digits, i.e. the information increases by one bit every time the number of different messages doubles.

\begin{figure}[ht]
\includegraphics[width=0.48\textwidth]{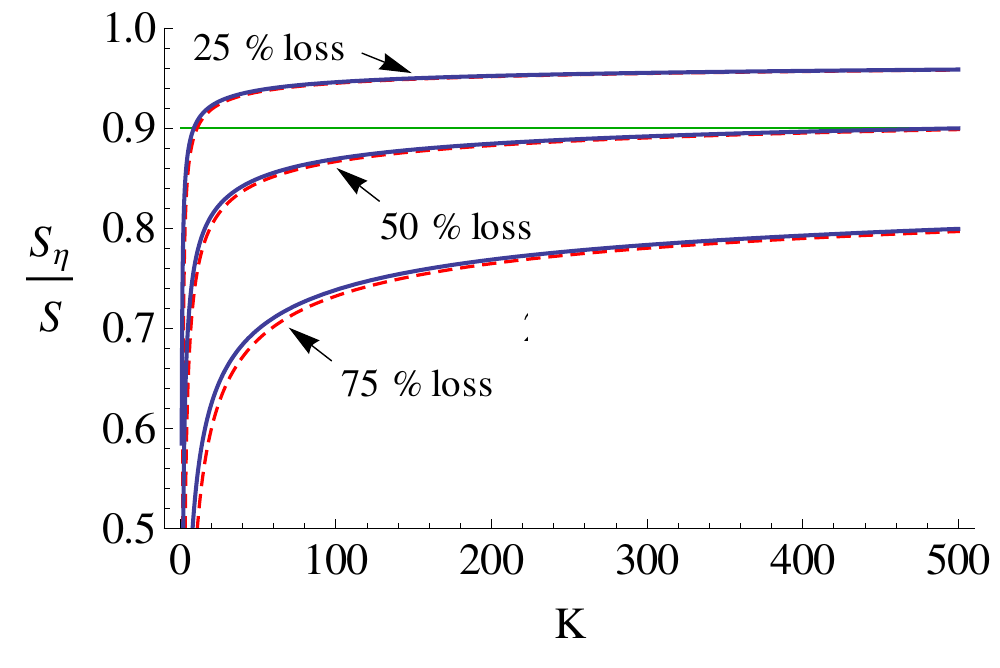}
\caption{\label{figfrac}A plot of the fraction of shared bits per photon pair as a function of $K$ for three types of measurements that yield different amounts of loss of joint modes. The solid lines are derived from the logarithmic approximation \eqref{vonneuman}, the dashed lines are derived from the non-approximated Von Neumann entropy.
}
\end{figure}

To illustrate this idea consider the following example.  Suppose that one can generate a state with a large Schmidt number, 
$K$, but that imperfections in the detection of the modes means that the number of effective modes that can be accessed is 
only $K/2$, i.e. $\eta=1/2$.  The number of shared bits will then be $S(K/2)=S(K)-1$: every time $K$ is halved, one shared bit is lost. As we show in figure \ref{figfrac}, the reduction in the entropy will be negligible for large enough values of $K$, i.e. for small (or large) enough values of $b\sigma$ (Fig. \ref{figfrac2}). 

If one is interested in determining the experimental parameters needed to retain a certain amount of shared bits, it is useful to recast Fig. \ref{figfrac} in terms of $b\sigma$:
\begin{figure}[ht]
\includegraphics[width=0.48\textwidth]{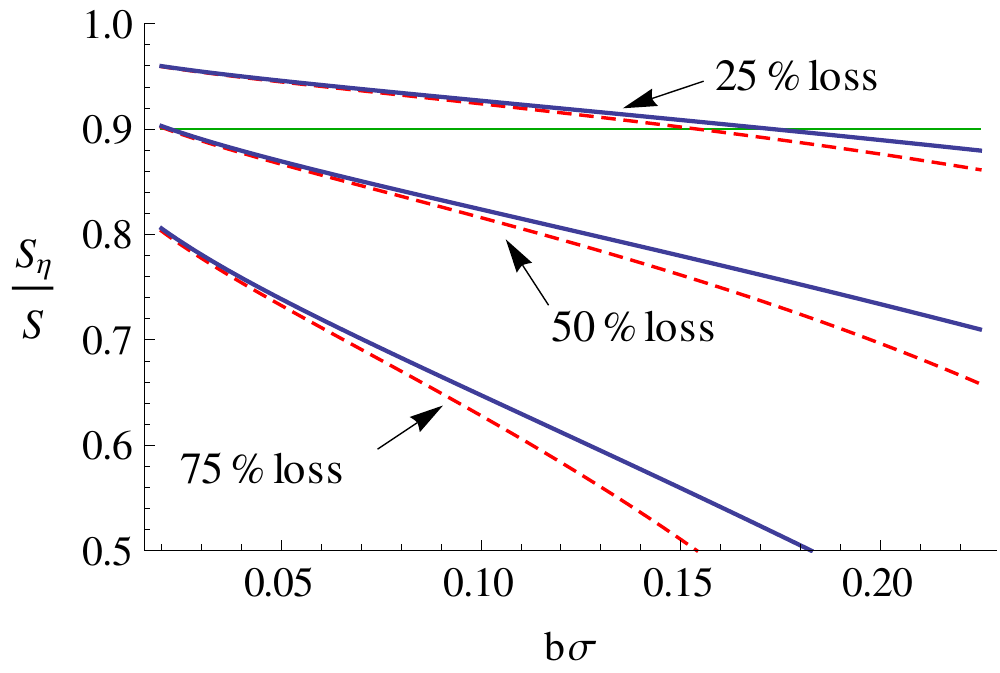}
\caption{\label{figfrac2}A plot of the fraction of shared bits per photon pair as a function of $b\sigma$ for three types of measurements that yield different amounts of loss of joint modes. The solid lines are derived from the logarithmic approximation \eqref{vonneuman}, the dashed lines are derived from the non-approximated Von Neumann entropy.
}
\end{figure}

To give some realistic numbers, even if the detection basis allows only half of the modes to be detected, it is still possible to retain more than 90\% of the shared bits, if states with more than \mbox{$\sim500$} entangled modes are used, which corresponds to $b\sigma\sim0.02$. Values of hundreds of entangled modes can be produced within the limitations of experimental equipment \cite{Pires09}.

\section{Conclusions}
We have derived analytical Schmidt decompositions for the biphoton state produced using SPDC in both cartesian and polar
coordinates for cases when the sinc phase matching term can be approximated as a gaussian. 
The resultant modes exhibit either the orthogonal characteristics of Hermite-Gauss modes of width $\Gamma$ or the angular and 
radial characteristics of Laguerre-Gauss modes of width $\Gamma/\sqrt2$, respectively, and can be shown to be equivalent using the relation between HG and LG modes. An entropic analysis of these different derivations confirms that the strength of the spatial entanglement is independent of the choice of coordinates and gives the freedom to choose the Schmidt basis that is most appropriate for given experimental conditions.

In order to analyse the entanglement we calculated the Schmidt number and the R\'enyi  and von Neumann entropies for the SPDC biphoton.
We showed under what conditions it is safe to use a logarithmic relation between the number of Schmidt modes and the maximum number of bits per photon that one can extract.
Such logarithmic relation demonstrates that, for highly entangled states, the information loss due to non-ideal measurements can be only a small fraction of the maximum information.
We finally showed, given any non-ideal measurement, which will allow to detect a smaller number of entangled modes $K_{\mathrm{eff}}$, what is the Schimdt number $K$ that is needed in order to retain a given fraction of shared bits, and from this what is the experimental parameter $b\sigma$ that one should look for.

\appendix
\section{Cartesian decomposition} \label{sec-AppA}
To perform the decomposition in cartesian coordinates we require the following mathematical result
\begin{align}
\label{main}
e^{-G(x^2+y^2-2\eta xy)}=\sqrt{1-|\mu|^2}\sum_{n=0}^{\infty}{\mu^nh_n(\Gamma x)h_n(\Gamma y)}
\end{align}
for $|\eta|<1$, $G>0$ and where 
\begin{align}
h_n(\Gamma w)=\frac{\sqrt{\Gamma}e^{-\Gamma^2w^2/2}H_n(\Gamma w)}{(n!2^n\sqrt{\pi})^{1/2}},\nonumber
\end{align}
where $H_n(v)$ are Hermite polynomials. A proof of this formula can be easily obtained with the use of generating functions for the Hermite polynomials.

In order to apply this result to the state \eqref{spdcstate} we set $G=b^2+1/\sigma^2$ and $\eta=(b^2\sigma^2-1)/(b^2\sigma^2+1)$. A bit of algebra gives the relations 
\begin{align}
\mu=\left|\frac{G\eta}{(G+\Gamma^2/2)}\right|=\left|\frac{b\sigma-1}{b\sigma+1}\right| \, ;  \Gamma=\sqrt{\frac{4b}{\sigma}} .
\end{align}

Equation \eqref{spdcstate} can thus be written in the form

\begin{align}
\psi&=\mathcal{N} e^{-G(q^2_i+q^2_s-2\eta q_iq_s)}e^{-G(q^2_{i\bot}+q^2_{s\bot}-2\eta q_{i\bot}q_{s\bot})}\nonumber\\
&=(1-\mu^2)\sum_{m,n}{\mu^m\mu^n h_m(\Gamma q_i)h_m(\Gamma q_s) h_n(\Gamma q_{i\bot}) h_n(\Gamma q_{s\bot})}
\end{align}
Let $h_{mn}(x,y)=h_m(\Gamma x)h_n(\Gamma y)$.  Using the properties of Hermite polynomials one can verify that these functions form a complete orthonormal set for $\mathcal{L}^2(\mathbb{R}^2)$ and the result in eq. \eqref{schm} follows.

\section{Polar decomposition} \label{sec-AppB}
Unlike for the cartesian decomposition which was performed in one step because cartesian orthogonal degrees of freedom 
play the same role, for the polar decomposition it is necessary to separate the angular variables and the radial variables 
in a different way. The angular variables will be separated with the Fourier transform, while the radial variables will be separated 
with a variation of formula \eqref{main}.

As the first step, we can rewrite the wave function \eqref{spdcstate} in polar coordinates and show that it is a function of the 
\emph{difference} of the angular variables. This fact enforces the conservation of OAM and allows to write it as a sum over the 
Fourier components of the difference of the angular variables:
\begin{align}
\label{state}
\psi=\mathcal{N}\exp\bigg[&-\frac{1}{\sigma^2}\left(\rho_i^2+\rho_s^2+2\rho_i\rho_s\cos(\varphi_i-\varphi_s)\right)\nonumber\\
&-b^2\left(\rho_i^2+\rho_s^2-2\rho_i\rho_s\cos(\varphi_i-\varphi_s)\right)\bigg]\\
&\hspace{-1.7cm}=\frac{1}{2\pi}\sum_\ell\sqrt{P_\ell}F_\ell(\rho_i,\rho_s)e^{i\ell(\varphi_i-\varphi_s)} 
\end{align}
where the sum runs over all integers.
The Fourier components are easily found:
\begin{align}
\sqrt{P_{\ell}}F_{\ell}(\rho_i,\rho_s)=\mathcal{N}e^{-\left(b^2+\frac{1}{\sigma^2}\right)\left(\rho_i^2+\rho_s^2\right)} I_{|\ell|}\left[2\left(b^2-\frac{1}{\sigma^2}\right)\rho_i\rho_s\right]
\label{FT}
\end{align}
Where $I_\ell(\cdot)$ is the $\ell^{\mathrm{th}}$ order modified Bessel function of the first kind.

The next step is to decompose each angular eigenfunction into a radial superposition of orthogonal modes.
The mathematical result needed to proceed is:
\begin{align}
&\sum_{p=0}^\infty \mu^{2p}r_p^{(\ell)}(x)r_p^{(\ell)}(y)=\frac{ |\mu|^{-|\ell|}}{1-\mu^2}e^{-\frac{x^2+y^2}{2}\frac{1+\mu^2}{1-\mu^2}}I_{|\ell|}\left(2xy\frac{|\mu|}{\mu^2-1}\right)
\label{kernelHH}
\end{align}
a proof of which can be found in \cite{watsonHH}. Here the $r_p^{(\ell)}$ functions are given by
\begin{align}
r_p^{(\ell)}(x)=\sqrt{\frac{2p!}{(p+|\ell|)!}}e^{-\frac{x^2}{2}}x^{|\ell|}L_p^{(|\ell|)}(x^2)
\end{align}
where $L_p^{(\ell)}$ are generalized Laguerre polynomials.

We apply the formula \eqref{kernelHH} to the functions in eq. \eqref{FT}. A  bit of algebra yields the correct value of the parameter $\mu$ and the correct scaling, $\Gamma$, of the $r_p^{(\ell)}\left(\Gamma{\rho}\right)$ functions:
\begin{align}
\label{eigenvalue}
\mu^2=\left(\frac{1-b \sigma}{1+b \sigma}\right)^2 \, ; \Gamma=\sqrt{\frac{4b}{\sigma}} .
\end{align}
Notice that the values are analogous to the cartesian case.

Applying these results and normalizing the radial modes, the result in eq. \eqref{pol} follows.

\section*{Acknowledgements}
We thank Stephen Barnett for useful discussions.
This work was supported by the UK EPSRC.
We acknowledge the financial support of the Future and Emerging
Technologies (FET) program within the Seventh Framework Programme
for Research of the European Commission, under the FET Open grant
agreement HIDEAS number FP7-ICT-221906.
This research was supported by the DARPA InPho program through the US Army Research Office award W911NF-10-1-0395.

\bibliographystyle{epj}

\begin{thebibliography}{100}
\bibitem{qkd} A. K. Ekert, {Phys. Rev. Lett.} {\bf 67}, 661 (1991).
\bibitem{qkd2} A. K. Ekert, J. G. Rarity, P. R. Tapster and G. Massimo Palma, {Phys. Rev. Lett.} {\bf 69}, 1293 (1992).
\bibitem{qkd3} W. Tittel, J. Brendel, H. Zbinden and N. Gisin, {Phys. Rev. Lett.} {\bf 84}, 4737 (2000).
\bibitem{hyper1} P. G. Kwiat, {J. Mod. Opt.} {\bf 44}, 2173 (1997).
\bibitem{hyper2} J. T. Barreiro, N. K. Langford, N. A. Peters and P. G. Kwiat, {Phys. Rev. Lett.} {\bf 95}, 260501 (2005).
\bibitem{Jack10} B. Jack, A. M. Yao, J. Leach, J. Romero, S. Franke-Arnold, D. G. Ireland, S. M. Barnett, and M. J. Padgett, {Phys. Rev. A} {\bf 81}, 043844 (2010).
\bibitem{BoydNLOptics} R. W. Boyd, Nonlinear Optics (Academic Press, 2008).
\bibitem{Torres03} J. P. Torres, A. Alexandrescu and Lluis Torner {Phys. Rev. A} {\bf 68}, 050301 (2003).
\bibitem{Miatto11} F. M. Miatto, A. M. Yao and S. M. Barnett, {Phys. Rev. A}, {\bf 83}, 033816 (2011).
\bibitem{Yao11} A. M. Yao, {New J. Phys.} {\bf 13}, 053048 (2011).
\bibitem{barnett} S. M. Barnett, \textit{Quantum Information}, (Oxford University Press, Oxford, 2009).
\bibitem{indexcor}  S. M. Barnett and S. J. D. Phoenix {Phys. Rev. A.} {\bf 40}, 2404 (1989).
\bibitem{indexcor2} S. M. Barnett and S. J. D. Phoenix {Phys. Rev. A} {\bf 44}, 535 (1991).
\bibitem{hall} M. J. W. Hall, {Phys. Rev. A} {\bf 55}, 100 (1997).
\bibitem{hallcorr} M. J. W. Hall, E. Andersson and T. Brougham, Phys Rev. A {\bf 74}, 062308 (2006).
\bibitem {Hong85} C. K. Hong and L. Mandel, {Phys. Rev. A} {\bf 31}, 2409 (1985).  
\bibitem{carlosgauss} C. W. Monken, P. H. SoutoRibeiro and S. Padua {Phys. Rev. A} {\bf 57}, 3123 (1998).
\bibitem{laweberly} C. K. Law and J. H. Eberly, {Phys. Rev. Lett.} {\bf 92}, 127903 (2004).
\bibitem{nch} M. A. Nielsen and I. L. Chuang, \textit{Quantum Computation and Quantum Information}, (Cambridge University Press, Cambridge, 2000).
\bibitem{lawschmidt} C. K. Law, I. A. Walmsley and J. H. Eberly, {Phys. Rev. Lett.} {\bf 84}, 5304 (2000).
\bibitem{ekart} A. Ekert and P. L.  Knight, {Am. J. Phys.} {\bf 63}, 415 (1995).
\bibitem{hgtolg}E. Abramochkin and V. Volostnikov, Opt. Comm. {\bf83}, 123 (1991)
\bibitem{Straupe11} S. S. Straupe, D. P. Ivanov, A. A. Kalinkin, I. B. Bobrov and S. P. Kulik, {Phys. Rev. A} {\bf 83}, 060302 (2011).
\bibitem{Allen} L. Allen, M. W. Beijersbergen, R. J. C. Spreeuw and J. P. Woerdman, {Phys. Rev. A.} {\bf 45}, 8185 (1992).
\bibitem{shannon} C. E. Shannon and W. Weaver, The Mathemtiatical Theory of Communication, Urbana,
University of Illinois Press, 1949.
\bibitem{jaynes} E. T. Jaynes, {Phys. Rev.} {\bf 106}, 620 (1957).
\bibitem{EIT} T. M. Cover and J. A. Thomas, Elements of Information Theory, (John Wiley and Sons, 1991).
\bibitem{renyi} A. R\'eyni,  Proceedings of the 4th Berkeley Symposium on Mathematics, Statistics and Probability,  pp. 547 (1960).
\bibitem{FlammiaPRL09} S. T. Flammia, A. Hamma, T. L. Hughes and X. G. Wen {Phys. Rev. Lett.} {\bf 103}, 261601 (2009).
\bibitem{BDSW} C. H. Bennett, D. P. DiVincenzo, J.A. Smolin and W. K. Wootters, {Phys. Rev. A} {\bf 54}, 3824 (1996).
\bibitem{BBPS} C. H. Bennett, H. J. Bernstein, S. Popescu and B. Schumacher, {Phys. Rev. A} {\bf 53}, 2046 (1996).
\bibitem{FedorovSpace} M. V. Fedorov, M. A. Efremov, P. A. Volkov, E. V. Moreva, S. S. Straupe and S. P. Kulik, {Phys. Rev. A} {\bf 77}, 032336 (2008).
\bibitem{FedorovTime} Y. M. Mikhailova, P. A. Volkov and M. V. Fedorov, {Phys. Rev A} {\bf 78}, 062327 (2008).
\bibitem{ThomasPRA2012} T. Brougham and S. M. Barnett, {Phys. Rev. A} {\bf 85}, 032322 (2012).
\bibitem{Ma2007}X. Ma, C.-H. F. Fung, H.-K. Lo, {Phys. Rev. A} {\bf 76}, 012307 (2007). 
\bibitem{Pires09} H. Di Lorenzo Pires, C. H. Monken and M. P. van Exter, {Phys. Rev. A} {\bf 80}, 022307 (2009)
\bibitem{watsonHH}G. N. Watson, J. London Math. Soc. 8, 189 (1933).

\end{thebibliography}

\end{document}